# Resting state fMRI-based brain information flow mapping

Abbreviated title: brain information flow mapping

Ze Wang, PhD

ORCID: 0000-0002-8339-5567

Department of Diagnostic Radiology and Nuclear Medicine

University of Maryland School of Medicine

670 W. Baltimore St, Baltimore, MD 20201

ze.wang@som.umaryland.edu

Conflict of interest: none

Data availability statement: the HCP data is freely available from the HCP consortium.

Ethics approval statement: data reanalysis has been approved by IRB. Patient consent forms were obtained by HCP.

**Acknowledgements**

The research effort involved in this study was supported by NIH grants: R01AG060054, R01 AG070227, R01EB031080, R01AG081693, R21AG08243, R21AG080518, 5P41EB029460, and by the support of the University of Maryland, Baltimore, Institute for Clinical & Translational Research (ICTR) and the National Center for Advancing Translational Sciences (NCATS) Clinical Translational Science Award (CTSA) grant number 1UL1TR003098. Both imaging and behavior data were provided by the Human Connectome Project, WU-Minn Consortium (Principal Investigators: David Van Essen and Kamil Ugurbil; 1U54MH091657) funded by the 16 NIH Institutes and Centers that support the NIH Blueprint for Neuroscience Research; and by the McDonnell Center for Systems Neuroscience at Washington University in St. Louis. The author thanks the Human Connectome Project for open access to its data.

Abstract

Human brain is a massive information generation and processing machine. Studying the information flow may provide unique insight into brain function and brain diseases. We present here a tool for mapping the regional information flow in the entire brain using fMRI. Using the tool, we can estimate the information flow from a single region to the rest of the brain, between different regions, between different days, or between different individuals' brain.

**Introduction**

The human brain is a self-organized system(Haken, 2012; Singer, 2009; Willshaw, 2006) which relies on massive information generation and processing to coordinate and adapt its functions. Understanding the persistent information transmission flow across regions or across time or even imaginarily across individual brain will provide a critical window to look inside brain functions and disease alterations.

We have previously mapped regional brain entropy using resting state fMRI (Wang et al., 2014) which is related to the information of local brain activity. The so-called brain entropy (BEN) mapping has been shown to be informative of brain health and neurocognitions (Da Chang, 2018; Donghui Song, 2019; Liu et al., 2020; Wang, 2020; Wang and Initiative, 2020; Xue et al., 2019; Zhou et al., 2016). While it can be extended into the cross-regional entropy mapping as we recently shown, it does not characterize the directional information flow.

In this work, we used transfer entropy (TE)(Schreiber, 2000) to map the inter-regional or inter-sessional or inter-subject information flow. We implemented several TE calculation algorithms in C++ and parallel threading and then applied it to data from the Human Connectome Project (HCP) (Van Essen et al., 2013).

**Materials and Methods**

*Ethics statement*

Data acquisition and sharing have been approved by the HCP parent IRB. Written informed consent forms have been obtained from all subjects before any experiments. This study re-analyzed the HCP data and data Use Terms have been signed and approved by the WU-Minn HCP Consortium.

*Data included*

rsfMRI data, demographic data, and neurobehavior data from 1102 healthy young subjects were downloaded from HCP. After excluding subjects who did not have full rsfMRI scans, or demographic, or behavioral, 865 remained (age 22-37 yrs, male/female=401/464). The range of education years was 11-17 yrs with a mean and standard deviation of 14.86±1.82 yrs. The rsfMRI data used in this paper were the extended processed version released on July 21 2017. Each subject had four rsfMRI scans acquired with the same multi-band sequence(Moeller et al., 2010) but the readout directions differed: readout was from left to right (LR) for the 1$^{st}$ and 3$^{rd}$ scans and right to left (RL) for the other two scans. The purpose of acquiring different scans with opposite phase encoding directions was to compensate for the long scan time induced image distortion. MR scanners all present field strength (B0) inhomogeneity, which causes signal distortion because of the imperfect excitation using the radiofrequency pulses that are tuned to the frequency determined by the ideal B0. While the B0 inhomogeneity caused distortions can be well corrected using two additionally acquired calibration scans using the opposite phase encoding directions: one is with LR and the other is with RL, HCP acquired two LR and two RL rsfMRI scans for the purpose of assessing the potential residual effects after the distortion correction and to assess the test-retest stability of rsfMRI measure. Each scan had 1200 timepoints. Other acquisition parameters for rsfMRI were: repetition time (TR)=720 ms, echo time=33.1ms, resolution 2x2x2 mm$^3$. The pre-processed rsfMRI data in the Montreal Neurological Institute (MNI) brain atlas space were downloaded from HCP (the S1200 release) and were smoothed with a Gaussian filter with full-width-at-half-maximum = 6mm to suppress the residual inter-subject brain structural difference after brain normalization and artifacts in rsfMRI data introduced by brain normalization. Non-neural spatiotemporal signal components were removed using the ICA-FIX algorithm(Griffanti et al., 2014; Salimi-Khorshidi et al., 2014; Smith et al., 2013). Motion parameters and their derivatives were regressed out from the time series too. More preprocessing details can be found in the HCP data release manual.

*TE calculation*

Denote two time series, for example, the time series of two brain voxels, by $X = [x_1, x_2, ... x_N]$ and $Y = [y_1, y_2, ... y_N]$, where N is the total number of timepoints which can be different between X and Y though the same length was used here for the simplicity of description. TE(Schreiber, 2000) quantifies the influence of the past of one variable on the future status of the other or alternatively whether adding the previous status of one variable can better predict the current or future status of the other variable than simply relying on the past status of that variable itself or the dependence of future state of one variable on the past states of itself and another variable. Mathematically, TE is measured through the deviation from the generalized Markov property, which is $p(x_{i+1}|X_i^k, Y_i^l) = p(x_{i+1}|X_i^k)$ for TE from y to x (TE(Y->X) or simply TE(Y,X)). k and l are the corresponding time lags for x and y, respectively and are often set to be 1 to reduce the computation complexity. $X_i^k = (x_{i-k-1}, ..., x_i)$ and $Y_i^l = (y_{i-l-1}, ..., y_i)$. The divergence can be measured through the Kullback entropy(Cover, 1999) concept:

$$T(Y,X) = \sum p(x_{i+1}, X_i^k, Y_i^l) \log \frac{p(x_{i+1}|X_i^k, Y_i^l)}{p(x_{i+1}|X_i^k)} \quad (1)$$

Kullback entropy measures the error when an approximate probability distribution q(.) is used in calculating the Shannon entropy instead of the true probability distribution p(.). Suppose Shannon entropy of X is defined by $H(X) = \sum p(x_i) \log \frac{1}{p(x_i)}$ and q(.) is used in practice instead of p(.). Kullback entropy can be then obtained by averaging the weighted difference in uncertainty log(1/q(xi)) – log(1/p(xi)):

$$K_{p|q}(X) = \sum p(x_i) \log \frac{p(x_i)}{q(x_i)} \qquad (2)$$

To measure TE(Y,X), we can first quantify the information transition from the past to the future for the same process X, which is the conditional Shannon entropy(Cover, 1999):

$$H(x_{i+1}|X_i^k) = \sum p(x_i, X_i^k) \log \frac{1}{p(x_i, X_i^k)} \qquad (3)$$

Similar to Shannon entropy, we often do not know the true transition probabilities $p(x_i, X_i^k)$. When a priori ones $q(x_i, X_i^k)$ are used, the measurement error can be described using the Kullback entropy concept and leads to the conditional Kullback entropy(Cover, 1999; Kaiser and Schreiber, 2002):

$$K_{p|q}(x_{i+1}|X_i^k) = \sum p(x_i, X_i^k) \log \frac{p(x_i, X_i^k)}{q(x_i, X_i^k)} \qquad (4)$$

In TE(Y,X), we care about whether adding the past of Y changes the probability of using the past of X to predict future states of X. In other words, we care about the difference between $p(x_{i+1}|X_i^k, Y_i^l)$ and $p(x_{i+1}|X_i^k)$. This problem can be reexpressed as measuring the error of using $p(x_{i+1}|X_i^k)$ as an alternative to $p(x_{i+1}|X_i^k, Y_i^l)$ in the format of Shannon entropy, which is the joint probability weighted sum of the difference $\log\left(\frac{1}{p(x_{i+1}|X_i^k)}\right) - \log\left(\frac{1}{p(x_{i+1}|X_i^k, Y_i^l)}\right)$ through Equation 1:

Comparing Equation 1 to the conditional Kullback entropy given in Equation 4, we can find that TE(Y,X) is the conditional Kullback entropy when the transitional probability of X $p(x_{i+1}|X_i^k)$ is used as the alternative to the true transitional probability of $p(x_{i+1}|X_i^k, Y_i^l)$.

TE(X,Y) can be similarly defined as:

$$T(X,Y) = \sum p(y_{i+1}, Y_i^k, X_i^l) \log \frac{p(y_{i+1}|Y_i^k, X_i^l)}{p(y_{i+1}|Y_i^k)} \quad (5)$$

Similar to Shannon entropy, TE depends on an accurate estimation of the probability density function (PDF). Since its inception, TE has been estimated using a few different methods through different ways of probability density estimation (Kaiser and Schreiber, 2002; Schreiber, 2000; Staniek and Lehnertz, 2008). We have implemented the standard histogram PDF estimation based approach with or without using the symbolization (Staniek and Lehnertz, 2008), the Darbellay-Vajda partitioning(Darbellay and Vajda, 1999) PDF estimation based approach, the kernel density estimation based PDF based TE calculation(Lee et al., 2012), and the simplified format(Kaiser and Schreiber, 2002).

*Experiments*

We calculated six different types of TE maps or matrix: the seed-based cross regional TE maps, a brain atlas-based cross-regional TE matrix, the brain atlas-based cross scan session TE matrix, the brain atlas-based cross brain TE matrix, the intra-brain TE density map, the voxel-wise cross scan session TE maps, the voxel-wise cross brain TE maps.

For the seed-based TE mapping, we included the following seeds: precuneus/posterior cingulate cortex (PCC), left amygdala, right amygdala, left dorso-lateral prefrontal cortex (DLPFC), right DLPFC, left hippocampus, right hippocampus, left insula, right insula, left inferior parietal sulcus (IPS), right IPS, left temporal parietal junction (TPJ), right TPJ, whole brain grey matter, and a control region in the deep white matter. The PCC seed was defined by Fox et al as the seed for characterizing the default mode network(Fox et al., 2005). The other seeds were based on the Pickatlas(Maldjian et al., 2003).

For the TE matrix, we used two different brain atlases: the Brainnetome Atlas(Fan et al., 2016), the Schaefer2018 parcellation (1000 parcellations)(Schaefer et al., 2018; Yeo et al., 2011).

*Test-retest stability*

Brain TE test-retest stability was assessed by the intra-class correlation (ICC) (Shrout and Fleiss, 1979) between the corresponding REST1 and REST2 TEs: TE (Y,X), TE(X,Y), and TE(Y,X)-TE(X,Y) separately.

*Statistical analyses on the biological and cognitive associations of brain TE.*

The following analyses were performed to the intra-brain TE maps or matrices. To find the potential biological or neuropsychological associations of brain TE, I performed several voxelwise regression analyses for each of the three TE values collected at the REST1 session (averaged across the first LR and the RL scans) and the REST2 session (averaged across the second LR and the RL scans), separately. Biological measures included age and sex. Cognitive capability was measured by the total cognitive function composite score (CogTotalComp_Unadj) in the NIH toolbox (http://www.nihtoolbox.org) that is derived by averaging the normalized scores of each of the included fluid and crystalized cognition measures and then deriving scale scores based on the new data distributions. Higher scores mean higher levels of cognitive functioning. I used the fully processed rsfMRI data provided by the HCP consortium and I have recently demonstrated that the full processing including the noise component removal step successfully suppressed nuisance effects related to physiological confounds such as respirational fluctuations and cardiac cycles(Del Mauro and Wang, 2023). To further control the residual respiratory and cardiac effects, I included the respiration rate (RR) and heart rate (HR) as two additional nuisance variables in all regression models. RR and HR were calculated from the corresponding record for each rsfMRI scan session. Signal sampling rate of those records was 400 Hz, which is roughly 288 times faster than the sampling rate of rsfMRI. Both time series were low pass filtered with a cutoff of 5 Hz and 10 Hz for the respiration and cardiac data, respectively. Local maxima were then detected using Matlab (Mathworks, Natick, Massachusetts, United States) function islocalmax (Matlab 2021b). The first derivative of the time stamps of the local maxima was calculated. The rmoutliers function of Matlab was used to remove outlier peak to peak time differences. The remaining time differences of each recording were averaged and considered to be the respiration cycle and heart beat cycle, respectively. Subjects were excluded from the following analyses if their final mean respiration cycle was longer than 10 secs or if the heart beat cycle was longer than 4 secs. RR and HR of the two scans of each scan session were averaged and paired with the corresponding TE each scan session (the mean of the LR and RL scans).

Two regression models were built. The first included sex, age, RR, and HR in the model and was used to investigate TE vs sex and age correlations. The second included sex, age, RR, HR, education years, and total cognitive score as covariate and was used to study the correlation between TE and the total cognition. Because total cognition score was collected on the same date as the second rsfMRI session, mean LR and RL TE of REST2 were used in the second regression model. The multiple regression model was built and estimated using Nilearn (https://nilearn.github.io/).

The voxelwise significance threshold for assessing each of the association analysis results was defined by $p<0.05$. Multiple comparison (across voxels) correction was performed with the family wise error theory (Nichols and Hayasaka, 2003) or false detection rate ($q<0.05$). Image and statistical results were displayed using Mricron (https://www.nitrc.org/projects/mricron) developed by Chris Rorden.

*Statistical analyses for the cross-session or cross-brain TE*

We calculated mean, standard deviation for the cross-session or cross brain TE matrix or maps. We also used the cross-brain TE as a distance to classify the individual brain, similar to the other imaging based disease subtyping work.

**Results**

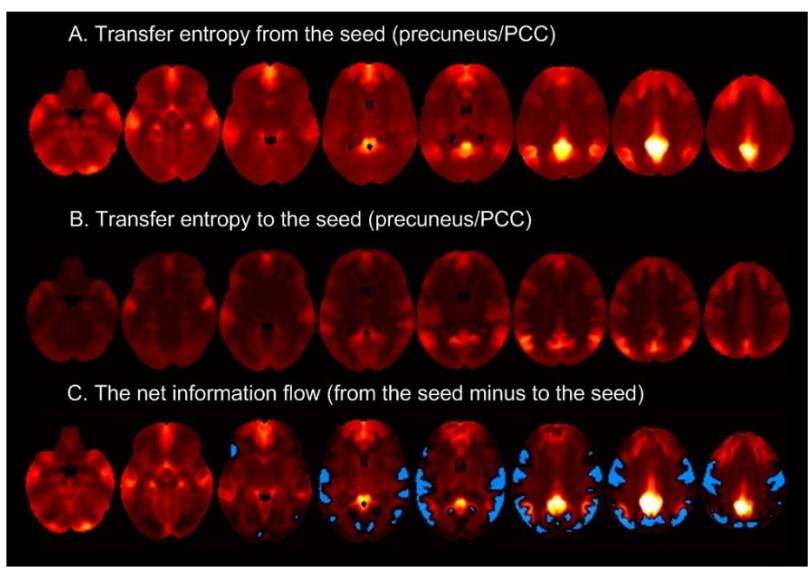

Fig. 1. Mean seed-based TE maps (n=1082). Seed was a sphere in precuneus/PCC. Red color means positive value. Blue means negative value.

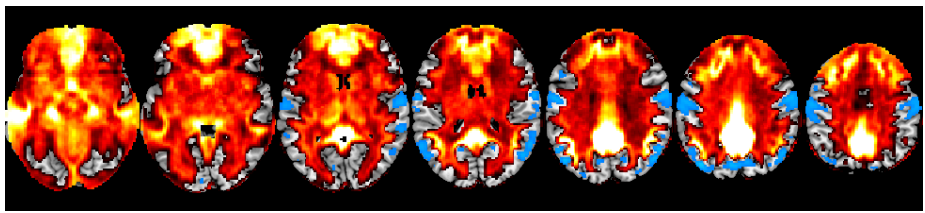

Fig. 2. One-sample t-test of the precuneus/PCC seed-based information flow (TE from the seed minus TE to the seed). Statistical significance level was defined by p<0.01 (FWE corrected). Hot

clusters mean that information flow in those regions is from the seed to those areas. Blue clusters mean that information flow in those regions is from them to the seed.

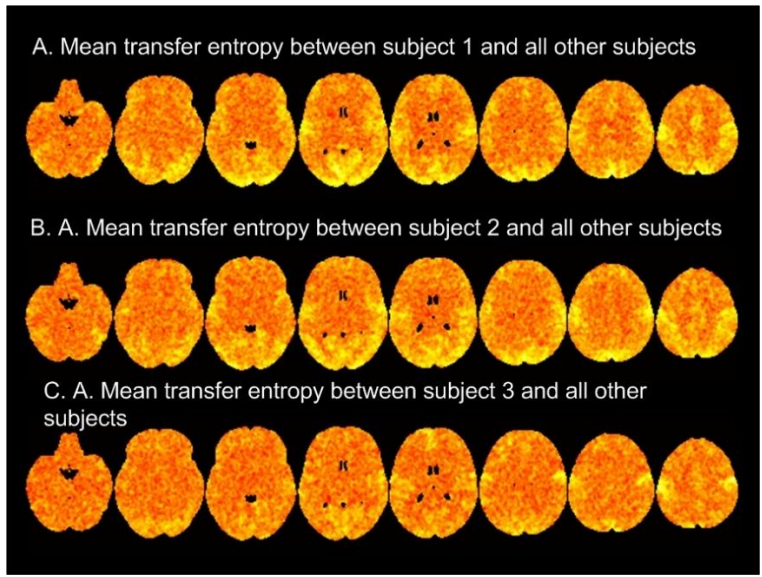

Fig. 3. Mean cross-subject TE of three representative subjects.

As mentioned in the above section, we applied TE mapping to the HCP data (Youth, Development, and Aging) from many different aspects. The computation was heavy and took several months to finish even using a high performance computing cluster from Institute for Genome Sciences at University of Maryland Baltimore. In this section, we started to show part of the second level analysis results. More will be added in future update of this preprint.

Fig 1 shows the mean precuneus/PCC seed-based TE maps. The top row is TE from the seed; the middle row is TE to the seed; the bottom row is information flow (top row – middle row). We can see clearly that information flows from dorso-lateral prefrontal cortex, middle temporal cortex, motor cortex, peripheral lateral parietal cortex, and peripheral superior precuneus to the precuneus/PCC and then to the inner lateral parietal cortex, temporal cortex, limbic system, and prefrontal cortex. Fig. 2 shows the family wise error corrected statistical inference results for the net information flow of the precuneus/PCC seed-based TE. The results were consistent with those shown in Fig. 1.

Fig. 3 shows the cross subject TE maps of three representative subjects. As it is difficult to generate a hypothesis about the directionally information flow between different individuals (except for several particular situations such as twins or lovers or individuals with strong bounds), we took the mean of the two TE values: from one subject to another and vice versa at each voxel and then average it across all the subjects. Specifically, for each subject, we considered that subject as the seed brain. TE at each voxel was computed between that voxel's time series from the seed brain and the same voxel (voxel at the same spatial location) in any of the other subjects. We then calculated the mean of the from-the-seed-brain TE and the to-the-seed-brain TE and averaged across all subjects for that seed brain. From Fig. 3, we can see that relatively high TE was found in the sensorimotor network (visual cortex, motor cortex) and the fronto-parietal network.

**Discussion**

This preprint describes a fMRI-based brain information flow mapping tool: BIF. We showed some initial results of secondary analyses of the many types of TE maps calculated from the HCP data. While the aim of this preprint was not to test a priori hypotheses, the results showed clear interesting patterns that can be used to generate hypotheses about within-brain and inter-brain information flow. The connectome-like TE matrices can be further used to investigator a potential information flow pathway which we will present results in the coming update. The cross-subject TE measures provide a way to assess the potential inter-subject brain activity distance. The code will be shared by request before we published the manuscript and will be through https://github.com/zewangnew.